\documentclass[twocolumn,10pt]{revtex4}

\usepackage{graphicx}
\usepackage{dcolumn}
\usepackage{bm}
\usepackage{color}
\usepackage{amssymb,amsmath}

\input epsf

\begin{document}

\title{\Large Thermodynamics of FRW Universe: Heat Engine}

\author{\bf Ujjal Debnath\footnote{ujjaldebnath@gmail.com}}

\affiliation{Department of Mathematics, Indian Institute of
Engineering Science and Technology, Shibpur, Howrah-711 103,
India.\\}

\begin{abstract}
We assume the non-flat Friedmann-Robertson-Walker (FRW) Universe
as a thermodynamical system. We assume the cosmological horizon as
a inner trapping horizon which is treated as dynamical apparent
horizon of FRW Universe. We write the dynamical apparent horizon
radius and temperature on the apparent horizon. We assume that the
fluid pressure as thermodynamical pressure of the system. Using
Hayward's unified first law, Clausius relation and Friedmann
equations with cosmological constant, we obtain the entropy on the
apparent horizon. We assume that the cosmological constant
provides the thermodynamic pressure of the system. We obtain the
entropy, surface area, volume, temperature, Gibb's Helmholtz's
free energies, specific heat capacity of the FRW Universe due to
the thermodynamic system. We study the Joule-Thomson expansion of
the FRW Universe and by evaluating the positive sign of
Joule-Thomson coefficient, we determine that the FRW Universe
obeys the cooling nature. We also find the inversion temperature
and inversion pressure. Next we demonstrate the thermodynamical
FRW Universe as heat engine. For Carnot cycle, we obtain the work
done and the maximum efficiency. Also for new engine, we study the
work done and its efficiency.\\
\\
Keywords: FRW Universe, Unified first law, Thermodynamics, Heat
Engine.
\end{abstract}

\maketitle


\section{Introduction}

Since the discovery of Hawking's radiation \cite{Haw1,Haw2}, the
black hole thermodynamics has become an intensive research topic
in Astrophysics. From the early discoveries of black hole
thermodynamics, it was speculated that the black hole area behaves
as thermodynamic entropy \cite{B,Bek2} and surface gravity behaves
as temperature \cite{Haw1}. Further, Hawking et al \cite{Hawking}
have analyzed the thermodynamic properties of Schwarzschild-AdS
black hole. In the study of black hole chemistry, the negative
cosmological constant (i.e., $\Lambda<0$) has been assumed as
thermodynamic pressure $P=-\frac{\Lambda}{8\pi}=\frac{3}{8\pi
\ell^{2}}$ where $\ell$ is the length of AdS black hole
\cite{K1,K2,Kubiz,K3,K4,K5}. The geometry of AdS black hole
thermodynamics has been studied by several authors
\cite{G1,G2,G3,G4,G5,G6,G7}. It is well known result that in
general relativity, the entropy to black hole horizon \cite{Haw1}
and cosmological horizon \cite{Gibb} have the same entropy-area
relation $S=A/4G$ where $A$ is the proper area of the horizon
surface. In the black hole thermodynamics, there are some
relations between black hole thermodynamics and Einstein's field
equations. Jacobson \cite{Jac} has observed that Einstein's field
equations can be derived from the relation
$S=A/4G$ on any local Rindler causal horizons.\\

Most discussions of black hole thermodynamics have been
concentrated on the stationary black holes. For non-stationary
(i.e., dynamical) black holes, Hayward \cite{Hay00,Hay0,Hay} has
proposed a mechanism of dynamical black hole thermodynamics
associated with its trapping horizon. Using this mechanism, for
spherical symmetric space-times, the Einstein's field equations
can be written in a form of Hayward's ``unified first law". From
this unified first law along trapping horizon, one can obtain the
first law of thermodynamics for dynamical black hole. In Hayward's
proposal, outer trapping horizon of dynamical black hole can be
used to the apparent horizon. Since the Friedmann-Robertson-Walker
(FRW) Universe is treated as one kind of non-stationary
spherically symmetric space-times, so similar to the study of
non-stationary black hole, we can discuss its thermodynamic
properties on the trapping horizon. In the FRW Universe, the outer
trapping horizon cannot exist, instead there exists a kind of
cosmological horizon like inner trapping horizon. So in the FRW
Cosmology, this horizon coincides with the apparent horizon. Cai
and Kim \cite{CaiKim} have obtained the Friedmann equations in
Einstein's gravity, Gauss-Bonnet gravity and Lovelock gravity
using the unified first law with the help of entropy on the
apparent horizon. Cai and Cao \cite{CC} have studied the unified
first law and thermodynamics of apparent horizon in FRW Universe
in the framework of Einstein theory, Lovelock theory and
scalar-tensor theory. Akbar and Cai \cite{Akbar,Akbar1,Akbar2}
have studied the thermodynamic phenomena of FRW Universe in
Einstein's gravity, scalar-tensor gravity, $f(R)$ gravity,
Gauss-Bonnet gravity and Lovelock gravity. From modified entropy
area relations, the modified Friedmann equations in FRW Universe
have been obtained by several authors
\cite{CCH1,Liu1,Yuan1,Shey1,Sar1,Giar1}. From these point of view
we'll study the thermodynamics in FRW Universe and its associated
thermodynamic quantities.\\

In the black hole thermodynamics, Johnson \cite{John} has proposed
the concept of holographic heat engine for AdS black hole, where
cosmological constant has been treated as a thermodynamic
variable. Subsequently, Johnson \cite{Joh,Joh1,Joh2,Johns} has
studied the Gauss-Bonnet black hole heat engine and Born-Infeld
AdS black hole heat engine. Based on the Johnson's holographic
heat engine proposal, several authors have studied the heat engine
phenomena and its efficiency for various types of AdS black holes
\cite{MoL,Hen,Liu,Mo1,Hendi,Pana,J,Avik,YeC,Gha,Johns11,Js}.
Recently, we have studied the thermodynamics, $P$-$V$ criticality,
stability, Joule-Thomson expansion and heat engine efficiency due
to Carnot cycle and Rankine cycle for AdS black holes
\cite{Por,Debnath1,Debnath2}. In the framework of the Universe,
the thermodynamics heat engine has been studied by Pilot
\cite{Pilot}. For polytropic gas model of the Universe, the Carnot
cycle heat engine phenomena has been studied by Askin et al
\cite{Askin}. Motivated by these works, here we'll study the
unified first law for FRW Universe and found the form of entropy
using Einstein's field equations and vice versa. Also we study the
thermodynamic quantities as well as Joule-Thomson expansion and
analyzed the heat engine phenomena for FRW Universe. In section
II, we study the Hayward's unified first law of thermodynamics for
FRW Universe. Using the Clausius relation, we derive the form of
entropy using Friedmann equations with cosmological constant. Then
in section III, we obtain the thermodynamic quantities and
Joule-Thomson expansion. In section IV, we study the phenomena of
heat engine for FRW Universe and study the Carnot engine and new
engine with work done and efficiencies. Finally, we present the
results of the whole work in section V.\\

\section{Unified First Law and Entropy of FRW Universe}

The line element for homogenous, isotropic and non-flat
Friedmann-Robertson-Walker (FRW) Universe is given by
\begin{equation}\label{FRW}
ds^2=-dt^2 +a^2(t)\left[\frac{dr^2}{1-kr^2}+r^2\left(d\theta^2 +
sin^2 \theta d\phi^2 \right)\right]
\end{equation}
where $a(t)$ is the scale factor and $k~(=0,-1,+1)$ represent the
flat, open and closed model of the Universe. We consider the
Einstein-Hilbert action as in the form
\begin{equation}\label{EH}
{\cal S}=\int d^4x\sqrt{-g}\left[\frac{R-2\Lambda}{16\pi
G}+{\cal{L}}_m\right]
\end{equation}
where $R$ is the Ricci scalar, $\Lambda$ is cosmological constant,
${\cal L}_{m}$ is the matter Lagrangian and $g=det(g_{ij})$
(choosing $c=1$). The Einstein's field equation is $G_{ij}=8\pi G
T_{ij}$. Here $T_{ij}$ is the energy momentum tensor for perfect
fluid, given by
\begin{equation}
T_{ij}=(\rho+p)u_{i}u_{j}+pg_{ij}
\end{equation}
where $\rho$ and $p$ are respectively the energy density and
pressure of perfect fluid. The four velocity $u_{i}$ satisfies the
relations $u_{i}u^{i}=-1$ and $u^{i}\nabla _{j}u_{i}=0$. The FRW
metric (\ref{FRW}) can be expressed in the following form
\cite{Bak}
\begin{equation}
ds^2=h_{ij}dx^{i}dx^{i}+\tilde{r}^2\left(d\theta^2 + sin^2 \theta
d\phi^2 \right)
\end{equation}
where $x^{0}=t$, $x^{1}=r$, $\tilde{r}=a(t)r$ and
$h_{ij}=diag(-1,~\frac{a^{2}}{1-kr^{2}})$. We consider the FRW
universe as a thermodynamical system and so the dynamical apparent
horizon, a marginally trapped surface with vanishing expansion can
be described by the relation $h^{ij}\frac{\partial
\tilde{r}}{\partial x^{i}}\frac{\partial \tilde{r}}{\partial
x^{j}}=0$. From this relation, we obtain the apparent horizon
radius $\tilde{r}_{h}$ for FRW Universe as in the form
\begin{equation}\label{AH}
\tilde{r}_{h}=\frac{1}{\sqrt{H^{2}+\frac{k}{a^{2}}}}
\end{equation}
where $H=\frac{\dot{a}}{a}$ is the Hubble parameter. Taking
derivative with respect to time $t$, we obtain
\begin{equation}\label{dAH}
\dot{\tilde{r}}_{h}=-\tilde{r}_{h}^{3}H\left(\dot{H}-\frac{k}{a^{2}}\right)
\end{equation}
Since the dynamical apparent horizon is the causal horizon
\cite{Bak,Hay,Hay1}, so it is associated with the gravitational
entropy and surface gravity. The surface gravity on the apparent
horizon is given by \cite{CaiKim}
\begin{equation}
\kappa=\frac{1}{2\sqrt{-h}}\frac{\partial}{\partial x^{i}}\left(
\sqrt{-h}~h^{ij}\frac{\partial \tilde{r}}{\partial x^{j}}\right)
\end{equation}
where $h=det(h_{ij})$. For FRW Universe, we obtain the surface
gravity on the apparent horizon as
\begin{equation}
\kappa=-\frac{1}{\tilde{r}_{h}}\left(1-\frac{\dot{\tilde{r}}_{h}}{2\tilde{r}_{h}H}
\right)
\end{equation}
So the temperature on the apparent horizon is obtained by
\begin{equation}\label{T1}
T=\frac{|\kappa|}{2\pi}=\frac{1}{2\pi\tilde{r}_{h}}\left(1-\frac{\dot{\tilde{r}}_{h}}{2\tilde{r}_{h}H}
\right)
\end{equation}
Within an infinitesimal time interval $dt$, if we assume that the
apparent horizon radius $\tilde{r}_{h}$ is kept fixed, then
$\dot{\tilde{r}}_{h}\ll 2\tilde{r}_{h}H$. In that approximate
case, there is no volume change in it and so we get
\cite{CaiKim,CCHu}
\begin{equation}\label{T2}
T= \frac{1}{2\pi\tilde{r}_{h}}
\end{equation}
which is identical with the Hawking temperature on the black hole horizon. \\

The Hayward's unified first law is defined by
\cite{Bak,Hay,Hay1,CC}
\begin{equation}\label{E1}
dE=A\Psi+WdV
\end{equation}
where $dE$ is the change of energy inside the apparent horizon.
The surface area $A$ of the apparent horizon is given by
\begin{equation}\label{A}
A=4\pi \tilde{r}_{h}^{2}
\end{equation}
and $V$ is the volume inside the apparent horizon surface, given
by
\begin{equation}\label{V0}
V=\frac{4\pi}{3} \tilde{r}_{h}^{3}
\end{equation}
The work done by change of the apparent horizon describes the work
density and the work density function $W$ is given by
\cite{Bak,Hay,Hay1,CC}
\begin{equation}
W=-\frac{1}{2}T^{ij}h_{ij}=\frac{1}{2}(\rho-p)
\end{equation}
The total energy flow through the apparent horizon denotes
energy-supply vector and is given by
\begin{equation}
\Psi_{i}=h^{jk}T_{ik}\frac{\partial \tilde{r}}{\partial
x^{j}}+W\frac{\partial \tilde{r}}{\partial x^{i}}
\end{equation}
From this, we obtain the components of energy-supply vector,
\begin{equation}
\Psi_{t}=-\frac{1}{2}(\rho+p)H\tilde{r}~,~\Psi_{r}=\frac{1}{2}(\rho+p)a.
\end{equation}
So we obtain the energy flux as in the form
\begin{equation}
\Psi=\Psi_{i}dx^{i}=-\frac{1}{2}(\rho+p)H\tilde{r}dt+\frac{1}{2}(\rho+p)adr
\end{equation}
Therefore the equation (\ref{E1}) yields
\begin{equation}\label{Psi1}
A\Psi+WdV=-A(\rho+p)H\tilde{r}dt+A\rho d\tilde{r}
\end{equation}
Since heat is one of the forms of energy, so according to the
study of Cai and Kim \cite{CaiKim}, the heat flow $\delta Q$
through the apparent horizon is just the amount of energy crossing
the apparent horizon during the time interval $dt$, i.e., $\delta
Q=-dE$. So using (\ref{E1}) and (\ref{Psi1}), on the apparent
horizon, we obtain
\begin{equation}\label{dQ}
\delta Q=-dE=A(\rho+p)H\tilde{r}_{h}dt
\end{equation}
The first law of thermodynamics (Clausius relation) on the
apparent horizon is
\begin{equation}\label{dQ1}
\delta Q=TdS
\end{equation}
From equations (\ref{dQ}) and (\ref{dQ1}), we obtain
\begin{equation}\label{TdS}
TdS=A(\rho+p)H\tilde{r}_{h}dt
\end{equation}

For the FRW metric, the Friedmann equations in presence of
cosmological constant in Einstein's gravity are given by
\begin{equation}\label{F1}
H^2+\frac{k}{a^2}=\frac{8\pi G}{3}~\rho+\frac{\Lambda}{3}
\end{equation}
and
\begin{equation}\label{F2}
\dot{H}-\frac{k}{a^2}=-4\pi G(\rho+p)
\end{equation}
We consider the Universe is filled with the fluid content whose
energy density and pressure are $\rho$ and $p$, satisfy the energy
conservation equation
\begin{equation}\label{Cons}
\dot{\rho}+3H(\rho+p)=0
\end{equation}
which can be obtained from the Einstein's field equations
(\ref{F1}) and (\ref{F2}). From equations (\ref{AH}) and
(\ref{F1}), we obtain
\begin{equation}\label{rho}
\rho=\frac{3}{8\pi G \tilde{r}_{h}^{2}}-\frac{\Lambda}{8\pi G}
\end{equation}
Also from equations (\ref{F1}) and (\ref{F2}), we obtain
\begin{equation}\label{p}
p=-\frac{1}{8\pi G \tilde{r}_{h}^{2}}+\frac{\Lambda}{8\pi G}
\end{equation}
Using the form of temperature (\ref{T2}), we now obtain the
entropy $S$ from the Friedmann equations (\ref{F1}) and
(\ref{F2}). Putting the expressions (\ref{T2}), (\ref{A}),
(\ref{rho}) and (\ref{p}) in the equation (\ref{TdS}) and then
integrating, we obtain the form of entropy as
\begin{equation}\label{SS}
S=\frac{A}{4G}+S_{0}
\end{equation}
where $S_{0}$ is an integration constant. If $S_{0}=0$, the above
entropy is identical form of entropy of the black hole horizon.
This is known result for FRW Universe in Einstein's gravity. \\

For FRW Universe, if we assume that the entropy has the form
$S=\frac{A}{4G}+S_{0}$ and temperature has the form (\ref{T2}),
then using conservation equation (\ref{Cons}) with the help of
(\ref{AH}), (\ref{dAH}) and (\ref{TdS}), we obtain the Friedmann
equations (\ref{F1}) and (\ref{F2}) which have also been found in
\cite{CaiKim}.

\section{Thermodynamic Quantities in FRW Universe}

For thermodynamic system, the entropy on the apparent horizon in
the FRW Universe is
\begin{equation}
S=\frac{A}{4G}+S_{0}=\frac{\pi
\tilde{r}_{h}^{2}}{G}+S_{0}~~\Rightarrow~
\tilde{r}_{h}=\sqrt{\frac{G}{\pi}}\sqrt{S-S_{0}}
\end{equation}
The volume inside apparent horizon can be written in terms of
entropy as
\begin{equation}\label{V}
V=\frac{4\pi}{3}~\tilde{r}_{h}^{3}=\frac{4}{3\sqrt{\pi}}~G^{3/2}(S-S_{0})^{3/2}
\end{equation}
The temperature (\ref{T2}) can be written in terms of entropy as
in the following form
\begin{equation}\label{TTT}
T=\frac{1}{2\sqrt{\pi G}\sqrt{S-S_{0}}}
\end{equation}
As well as in black hole thermodynamics, the cosmological constant
$\Lambda$ is teated as thermodynamic pressure $P$ i.e.,
$P=\frac{\Lambda}{8\pi G}$ and allows to variable. So from
equations (\ref{rho}) and (\ref{p}) we obtain
\begin{eqnarray}\label{rho1}
\rho=\frac{3}{8G^{2}(S-S_{0})}-P
\end{eqnarray}
and
\begin{eqnarray}\label{p1}
p=P-\frac{1}{8G^{2}(S-S_{0})}
\end{eqnarray}
We have taken positive $\Lambda$, otherwise $p$ will always
negative which cannot happen in general. Since $\rho>0$ so we have
$P<\frac{3}{8G^{2}(S-S_{0})}$ and hence $p<\frac{2P}{3}$. Also we
can write $\rho+3p=2P$. The enthalpy function is defined as ${\cal
H} = U + PV$ where $U$ is the energy. Hence, using the first law
of thermodynamics, we get
\begin{equation}
d{\cal H}=TdS+VdP
\end{equation}
Now integrating, we obtain the enthalpy function ${\cal H}$ as in
the following form,
\begin{equation}
{\cal H}=\frac{2\sqrt{S-S_{0}}}{3\sqrt{\pi G}}+
\frac{4}{3\sqrt{\pi}}~G^{3/2}\int (S-S_{0})^{3/2}dp+{\cal H}_{0}
\end{equation}
where ${\cal H}_{0}$ is integration constant. So the Gibb's free
energy can be obtained as \cite{Graca},
\begin{eqnarray}
&&{\cal G}={\cal H}-TS \nonumber
\\
&&=\frac{(S-4S_{0})}{6\sqrt{\pi G}~\sqrt{S-S_{0}}}+
\frac{4}{3\sqrt{\pi}}~G^{3/2}\int (S-S_{0})^{3/2}dp+{\cal H}_{0}
\nonumber
\\
\end{eqnarray}
Also the Helmholtz's free energy can be obtained as \cite{Graca}
\begin{eqnarray}
&&F={\cal G}-PV \nonumber
\\
&&=-\frac{S_{0}}{2\sqrt{\pi G}~\sqrt{S-S_{0}}}-
\frac{4}{3\sqrt{\pi}}~G^{3/2} (S-S_{0})^{3/2}p \nonumber
\\
&&+ \frac{4}{3\sqrt{\pi}}~G^{3/2}\int (S-S_{0})^{3/2}dp+{\cal
H}_{0}
\end{eqnarray}
The specific heat capacity of the FRW Universe in thermodynamical
system can be written as \cite{Kubiz}
\begin{eqnarray}
{\cal C}_{P}=T\left(\frac{\partial S}{\partial
T}\right)_{P}=2(S_{0}-S)
\end{eqnarray}
Coefficient of thermal expansion is given by
\begin{equation}
\alpha=\frac{1}{V}\left(\frac{\partial V}{\partial
T}\right)_{P}=-12\sqrt{\pi G}\sqrt{S-S_{0}}
\end{equation}
The isothermal compressibility is given by \cite{Cac}
\begin{equation}
\kappa_{T}=-\frac{1}{V}\left(\frac{\partial V}{\partial
P}\right)_{T}=\frac{3}{2(S-S_{0})}\left(\frac{1}{8G^{2}(S-S_{0})^{2}}-\frac{\partial
p}{\partial S} \right)^{-1}
\end{equation}
The minus sign accounts for the fact that an increase in pressure
generally induces a reduction in volume.\\

Joule-Thomson expansion \cite{Win,Jo} describes that the change of
temperature from high pressure regime to low pressure regime,
while the enthalpy remains constant. It is an irreversible
process, also known as the throttling process. Here, we now
examine the Joule-Thomson expansion for FRW Universe. The
Joule-Thomson coefficient $\mu$ is the slope of the isenthalpic
curve, defined by \cite{Ok1}
\begin{equation}
\mu=\left(\frac{\partial T}{\partial P}\right)_{{\cal H}}
\end{equation}
which can be written in the following forms
\begin{equation}\label{mu1}
\mu=\frac{1}{{\cal C}_{P}}\left[T\left(\frac{\partial V}{\partial
T}\right)_{P}-V \right]
\end{equation}
or
\begin{equation}\label{mu2}
\mu=\frac{1}{S}\left[P\left(\frac{\partial V}{\partial
P}\right)_{{\cal H}} +2V\right]
\end{equation}
The cooling or heating nature of the Universe can be determined by
the sign of $\mu$. If $\mu>0$, the cooling process occurs while
$\mu<0$ describes the heating nature. Now for FRW Universe, we
obtain
\begin{eqnarray}
\mu=\frac{8}{3\sqrt{\pi}}~G^{3/2}\sqrt{S-S_{0}}
\end{eqnarray}
We see that $\mu>0$ always, so cooling process occurs in FRW Universe.\\

Putting $\mu=0$ in (\ref{mu1}), we can obtain the expansion
process of inversion curve and so the inversion temperature is
obtained as
\begin{equation}
T_{inv}=V\left(\frac{\partial T}{\partial
V}\right)_{P}=-\frac{1}{12\sqrt{\pi G}\sqrt{S-S_{0}}}
\end{equation}
Also by putting $\mu=0$ in (\ref{mu2}), the inversion pressure can
be obtained as
\begin{equation}
P_{inv}=-2V\left(\frac{\partial P}{\partial V}\right)_{{\cal
H}}=\frac{4(S-S_{0})}{3}\left(\frac{1}{8G^{2}(S-S_{0})^{2}}-\frac{\partial
p}{\partial S} \right)
\end{equation}
The inversion pressure depends on the fluid pressure $p$ and
entropy $S$ of the system.

\section{\bf{Heat Engine for FRW Universe}}

Thermodynamically, heat engine is a physical system which converts
heat/thermal energy into mechanical energy for doing mechanical
work. So heat engine transfers heat from hot region, where part of
the heat transforms into physical works while remaining part of
the heat is moved to cold region. So the heat engine works in a
cyclic manner where the heat/thermal energy produces in one part
of the cycle, which can do work in another part of the cycle. In
this section, we'll study the Carnot cycle of the heat engine for FRW Universe.\\

In 1824, Carnot introduced a theoretical thermodynamic cycle,
known as Carnot cycle and corresponding classical heat engine is
known as Carnot heat engine. Now we assume, $T_{H}$ and $T_{C}$
are respectively the temperatures of hot and cold regions, which
consist of upper and lower isothermal processes. For Carnot heat
engine, Johnson \cite{John} has shown the $P$-$V$ diagram in a
closed path in order to calculate the work done by the heat
engine. In the diagram, the heat flows are produced from level 1
to level 2 along the upper isotherm process, given as
$Q_{H}=T_{H}\triangle S_{1\rightarrow 2}=T_{H}(S_{2}-S_{1})$ and
also the exhausted heat formed from level 3 to level 4 along lower
isothermal process, given as $ Q_{C}=T_{C}\triangle
S_{3\rightarrow 4}=T_{C}(S_{3}-S_{4})$. In FRW Universe, using
relation (\ref{V}), the entropies $S_{i}$'s are related to the
volumes $V_{i}$'s as in the following
\begin{equation}\label{37}
S_{i}=\frac{1}{G}\left(\frac{3\sqrt{\pi}}{4}\right)^{2/3}V_{i}^{2/3}+S_{0}~,~i=1,2,3,4,
\end{equation}
The total work done by the Carnot heat engine is $W=Q_{H}-Q_{C}$.
The efficiency of the Carnot heat engine is defined by the ratio
of total work done and so the amount of heat energy along the
upper isotherm process is given as $
\eta_{_{Car}}=\frac{W}{Q_{H}}=1-\frac{Q_{C}}{Q_{H}} $. In Carnot
cycle, we have the relations  $V_{4}=V_{1}$ and $V_{3}=V_{2}$. So
the maximum efficiency for Carnot cycle is obtained as $
(\eta_{_{Car}})_{_{max}}
=1-\frac{T_{C}}{T_{H}}=1-\sqrt{\frac{S_{H}}{S_{C}}} $. Since
$T_{H}>T_{C}$, so maximum efficiency for Carnot cycle always
satisfies $0<(\eta_{_{Car}})_{_{max}}<1$. \\

Johnson \cite{John} has also described a new engine which includes
two isobars and two isochores/adiabats, where the heat flows show
along top and bottom lines. The total work done along the isobars
is given in the form
\begin{eqnarray}
W=\triangle P_{4\rightarrow 1}~\triangle V_{1\rightarrow
2}=(P_{1}-P_{4})(V_{2}-V_{1})
\end{eqnarray}
The net inflow of heat along the upper isobar is given by
\begin{eqnarray}
Q_{H}&=&\int_{T_{1}}^{T_{2}} {\cal C}_{P}(P_{1},T)dT \nonumber
\\
&=&\frac{1}{2\sqrt{\pi
G}}~\left(\sqrt{S_{2}-S_{0}}-\sqrt{S_{1}-S_{0}} \right)
\end{eqnarray}
The net exhaust of heat from the lower isobar is given by
\begin{eqnarray}
Q_{C}&=&\int_{T_{3}}^{T_{4}} {\cal C}_{P}(P_{4},T)dT \nonumber
\\
&=&\frac{1}{2\sqrt{\pi
G}}~\left(\sqrt{S_{4}-S_{0}}-\sqrt{S_{3}-S_{0}} \right)
\end{eqnarray}
So the thermal efficiency for the new heat engine for FRW Universe
is given in the form
\begin{eqnarray}
&\eta_{_{New}}&=\frac{W}{Q_{H}}=\frac{(P_{1}-P_{4})(V_{2}-V_{1})}{Q_{H}}
\nonumber\\
&&=\left[\frac{8G^{2}}{3}~(p_{1}-p_{4})+
\frac{(S_{4}-S_{1})}{3(S_{1}-S_{0})(S_{4}-S_{0})}\right]~\nonumber
\\
&&\times
\left(S_{1}+S_{2}-2S_{0}+\sqrt{(S_{1}-S_{0})(S_{2}-S_{0})}\right)\nonumber\\
\end{eqnarray}
where $p_{1}$ and $p_{4}$ denote the fluid pressures at stages $1$
and $4$ respectively. We see that the efficiency of the new engine
$\eta_{_{New}}> 0$ always if $p_{1}> p_{4}$ and $S_{4}> S_{1}$.

\section{Discussions and Concluding Remarks}

We have assumed the non-flat Friedmann-Robertson-Walker (FRW)
Universe as a thermodynamical system. We have also considered the
cosmological horizon as a inner trapping horizon which is treated
as dynamical apparent horizon of FRW Universe. We have found the
dynamical apparent horizon radius and temperature on the apparent
horizon. We have assumed that the cosmological constant $\Lambda$
for the Universe as thermodynamical pressure $P$ of the system.
Using the Hayward's `unified first law', we have obtained the
change of the energy $dE$ inside the apparent horizon. The
Friedmann equations in presence of cosmological constant with the
help of conservation equation, we have obtained the entropy on the
apparent horizon as in the form $S=\frac{A}{4G}+S_{0}$. If
$S_{0}=0$ then the entropy is identical form on horizon entropy of
black hole. Conversely, from entropy-area relation, the Friedmann
equations with cosmological constant can be recovered
\cite{CaiKim}. Due to the thermodynamic system, we have obtained
the surface area, entropy, volume, temperature, Gibb's free
energy, Helmholtz's free energy and specific heat capacity of the
FRW Universe. We have examined the Joule-Thomson expansion of FRW
Universe and evaluated the Joule-Thomson coefficient $\mu$. The
sign of $\mu$ presents the key role for heating or cooling nature
of the Universe. We have found that $\mu$ is positive, so we may
concluded that in the thermodynamic system, the FRW Universe
produces the cooling nature. Putting $\mu=0$, we have obtained the
inversion temperature and inversion pressure. The inversion
pressure always depends on the nature of the fluid pressure as
well as entropy. Next we have demonstrated the thermodynamical FRW
Universe as heat engine. For Carnot cycle, we have obtained the
work done and its maximum efficiency which satisfies
$0<(\eta_{_{Car}})_{_{max}}<1$. Also we have found the work done
and its efficiency for a new engine, which satisfies
$\eta_{_{New}}> 0$ always if $p_{1}> p_{4}$ and $S_{4}> S_{1}$. \\

\end{document}